# To Assess the Impact of Smart Cities on Urbanization Patterns in the United States

by Wayne S. Singh, ws2398@nyu.edu

New York University, New York, NY 10012, USA

May 2024

Singh, Wayne ws2398@nyu.edu

**Abstract**

This paper investigates the nexus between smart city initiatives and the evolving urbanization trends in the United States, a pivotal concern as populations in the U.S. convert to urban living at a pace that leads global trends. The study is motivated by the pressing need to understand how innovations embedded within the smart city paradigm shape urban landscapes using a mixed-method approach. Using the principles of Urban Complexity Theory, four variables were identified as relevant to smart cities' influence on urbanization patterns: smart city technology, government policy, environmental sustainability, and socioeconomic factors. A web-based survey was conducted on residents in Manhattan, New York, and Capitol Hill, Seattle, for a sample size of (n=50) using a five-point Likert scale response. The study combined the quantitative survey and qualitative research to assess each contributing variable and uncover the significant factors of urban migration to smart cities. Key findings suggest a departure from traditional urban growth models to indicate that implemented smart city technologies correlate with population density shifts, land use diversification, and infrastructural dynamism. The findings infer that residents' preference for smart city living is influenced by efficient urban mobility, visible environmental sustainability, and personal socioeconomic improvement. The implications of this research extend to urban planners, policymakers, and employers. The study concludes that when decision-makers include the key influencing factors of urban smart city residents in short and long-term planning, cities are optimized to accommodate the growing population.

**TABLE OF CONTENTS**

**PURPOSE OF STUDY**

Although constantly advancing, information technology experienced a significant milestone in late 2022 with the monumental milestone in Artificial Intelligence (AI) and Machine Learning (ML) (T. Wu et al., 2023). The magnitude of its importance is realized when put into context. Since Alan Turing's "Artificial Intelligence" theory was coined in the mid-1950s, development into maturing the concept has been marginal at best (Muggleton, 2014). Current developments are, therefore, more than sixty years in the making.

Since then, AI has impacted practically every industry, from real estate to software development, augmenting repetitive tasks and improving its application with each successive iteration in the last two years (George & George, 2023). Concepts like smart cities, which already depend heavily on technology, have accelerated their growth, scale, and adoption due to lower barriers to AI accessibility (DeHart et al., 2021). However, the rapid scaling of smart cities helped, in part, by AI-assisted technology may threaten the sustainability of the urban ecosystem and the quality of life of its citizens. By identifying the factors that influence urbanization, specifically through the lens of smart cities, a hypothesis may be drawn on the risks contributing to urban planning, capacity management, service delivery, and the subsequent quality of life of residents of modern American metropolitans.

**PROBLEM STATEMENT**

Land use patterns for the United States reveal that as of 2023, eighty-three percent of the population reside in urban areas, an increase of sixty-four percent compared to 1950. By 2050, urbanization will represent eighty-nine percent of the population (*U.S. Cities Factsheet*, 2023). The rapid conversion to urban living has prompted cities to innovate solutions to address density, congestion, healthcare, and other social issues.

The smart city framework has gained popularity among governments and municipalities as an effective strategy to manage the growing urbanization trend of the past two decades (Kitchin, 2014). A smart city has many strategic and operational components. Still, it is generally characterized by its ability to leverage Information and Communication Technology (ICT) to displace traditional urban planning and deliverables and deliver services to citizens efficiently (Neirotti et al., 2014).

A smart city urbanization plan, however, is an aggressive form of land transformation that directly impacts biodiversity and ecological landscapes within a metropolitan area (J. Wu et al., 2011). Furthermore, due to data bias, introducing the digitization approach to urbanization may exacerbate social disparities and inequalities for some groups purporting a less-than-inclusive community (Wang et al., 2021). Today's economies of scale achieved through technology advancements accelerate smart city expansion plans (Wang et al., 2021). Therefore, assessing its effectiveness in fostering equitable and viable growth is essential for citizens and metropolitans.

**RESEARCH QUESTIONS**

1. Does smart city technology ($X_1$) impact urbanization patterns in the United States (Y)?
2. Do smart city government policies ($X_2$) influence urbanization patterns (Y) in the United States?

3. Do environmental conditions in smart cities ($X_3$) impact urbanization patterns (Y) in the United States?

4. Do socioeconomic ($X_4$) factors impact smart cities' urbanization patterns (Y) in the United States?

VARIABLES

**Y Variable (dependent variable)**

Urbanization Patterns in the United States (dependent variable)

*Urbanization patterns represent changes in population distribution, the expansion of urban areas, and the transformation of land use and infrastructure. Urbanization is a key metric as it directly reflects the consequences of technological advancement, policy shifts, environmental strategies, and socioeconomic dynamics in urban areas. The patterns of urbanization reveal how smart city initiatives might influence migration, city planning, and the development of urban spaces (M. Shahidehpour et al., 2018).*

**X Variables (independent variables)**

**$X_1$: Smart City Technology (Jeong & Park, 2019), (Popescu, 2015)**

*Smart city technology covers a range of digital innovations, from IoT (Internet of Things) devices to advanced data analytics, used to optimize city functions and drive economic growth. Within context, these technologies contribute to efficient city management and sustainable development. Considering this variable is crucial, as it can directly affect how urban areas evolve, potentially leading to efficient, more responsive urban environments.*

**$X_2$: Government Policy (Takahashi et al., 2021), (Ahani & Dadashpoor, 2021)**

*Policy and governance frameworks are the rules and actions by which urbanization, including smart city initiatives, are guided and implemented. Examining the role of policy in either facilitating or hindering the development of smart cities, emphasizing how governance frameworks affect the deployment and effectiveness of urban management. Including policy as a variable is important because it shapes the context within which smart cities and its supported technologies are deployed and can significantly influence the success and direction of urbanization.*

**$X_3$: Environmental Conditions (Huang et al., 2021), (Zhu et al., 2021)**

*Environmental influence in the context of smart cities includes sustainability practices and the impact of urban development on natural ecological resources. Analyzing how smart city designs and technologies aim to reduce environmental footprints and whether this is achievable and sustainable is essential for understanding their impact on urbanization trends.*

**$X_4$: Socioeconomic (Nathaniel, 2021), (Goel & Vishnoi, 2022)**

*Socioeconomic factors encompass the social and economic aspects that influence and are influenced by urbanization, such as income distribution, employment opportunities, education, health services, and, ultimately, the movement of people to or away from areas like smart cities. Considering these factors offers an understanding of the relationship between quality of life and the inclusiveness of urban growth. They can also indicate whether smart city developments lead to equitable and inclusive urbanization or contribute to social stratification.*

Singh, Wayne ws2398@nyu.edu

**HYPOTHESES STATEMENTS**

Research Question 1: Does smart city technology (X1) impact urbanization patterns in the United States (Y)?

$H_0$: Smart city technology ($X_1$) significantly impacts urbanization patterns (Y) in the United States.

$H_a$: Smart city technology (X1) does not impact urbanization patterns (Y) in the United States.

Research Question 2: Do smart city government policies ($X_2$) influence urbanization patterns (Y) in the United States?

$H_0$: Government Policy ($X_2$) significantly impacts the urbanization patterns (Y) of smart cities.

$H_a$: Government Policy ($X_2$) does not impact urbanization patterns (Y) of smart cities.

Research Question 3: Do environmental conditions in smart cities ($X_3$) impact urbanization patterns (Y) in the United States?

$H_0$: Environmental sustainability ($X_3$) significantly impacts the urbanization patterns (Y) of smart cities.

$H_a$: Environmental sustainability ($X_3$) does not impact the urbanization patterns (Y) of smart cities.

Research Question 4: Do socioeconomic ($X_4$) factors impact smart cities' urbanization patterns (Y) in the United States?

$H_0$: Socioeconomic factors ($X_4$) significantly impact the urbanization patterns (Y) of smart cities.

$H_a$: Socioeconomic factors ($X_4$) do not impact the urbanization patterns (Y) of smart cities.

## METHOD

Effectively addressing a research question is fundamental in explaining the topic. Therefore, an effective research design provides a systematic plan that comprehensively addresses the research questions and controls variance (Dulock, 1993). The study aimed to identify the influence of smart cities on the urbanization patterns of United States residents in New York City and Seattle by assessing the relationship between the independent ($X_{1234}$) and dependent (Y) variables. The researcher adopted a cross-sectional approach to the study, conducted over nine weeks.

The researcher aggregated the qualitative and quantitative data to measure the impact of (X) variables on urbanization patterns: smart city technology ($X_1$), government policy ($X_2$), environmental sustainability ($X_3$), and socioeconomic factors ($X_4$). Descriptive statistics, correlation and regression analysis, and Analysis of Variance (ANOVA) methods were used to interpret the quantitative data. Comparative and thematic analysis was performed on the qualitative data to identify trends and patterns between the dependent and independent variables.

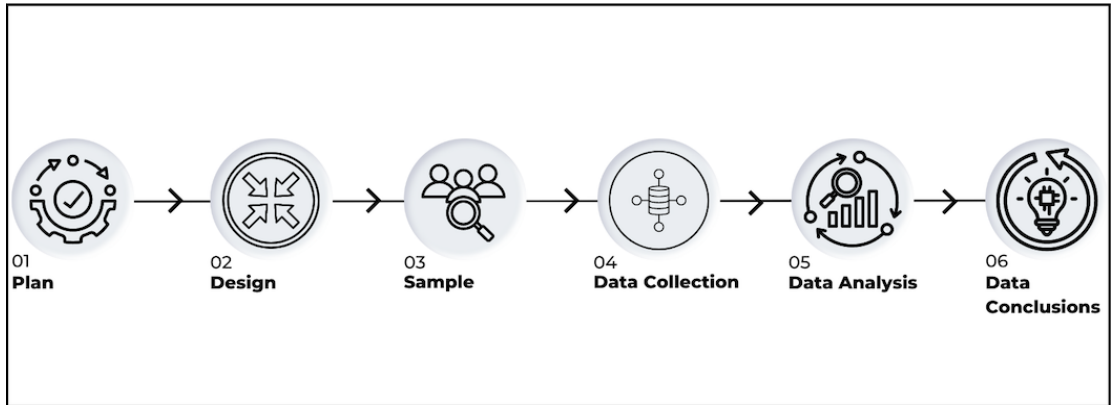

*Figure 1: Research phases for assessing the effects of smart cities on urbanization patterns in the United States*

Figure 1 is a graphical representation of the six-phase method utilized for the research design methodology. The variables were designed after the research plan was finalized. Sample scope, including demographic selections, was created in phase three. The data was then collected and analyzed for interpretation and conclusions during phases four to six.

The 'Plan' phase identified and established the research problems, subproblems, and questions to introduce complications to the topic. In addition, it provided input to determine the appropriate research methods for the 'Design' phase. The research framework and approach to address the qualitative or quantitative methods was decided in the 'Design' phase. The sample size (n=50) and scope were determined in the 'Sample' phase. The sample proportion for each city, New York and Seattle, was allocated during this phase, and the demographic requirements for each urban area were also established. The type and method of data collection for each survey category were defined in the 'Data Collection' phase, including the category of each survey group. The 'Data Analysis' phase was where appropriate analytical techniques were applied to the data collected. For example, descriptive statistics was used for qualitative data, and thematic analysis was used for quantitative data. The final phase, 'Data Conclusions', was where the interpretation of findings and possible conclusions were drawn between X variables and the Y variable for the research topic.

## VARIABLES

The researcher gathered information from academic research, including scholarly articles, to establish the four independent variables relevant to urbanization patterns in the United States. These variables are smart city technology, government policy, environmental sustainability, and socioeconomic factors, as depicted in Figure 2 below.

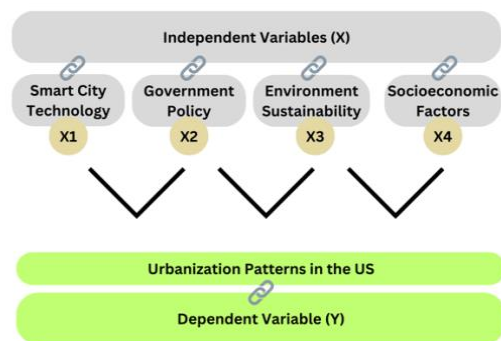

*Figure 2: Variable assessment categories*

The four independent variables selected had a substantial impact on urbanization patterns in the U.S. Firstly, the rapid growth of technology-supported infrastructure within the urban planning of a city and the subsequent services made available to residents were influencing factors that attracted people to urban, smart city life (Jeong & Park, 2019). Secondly, the success and viability of a smart city from an urban planner's perspective largely depend upon relevant state and federal policies (Ahani & Dadashpoor, 2021). Thirdly, for many municipal decision-makers, the primary incentive to adopt or embark on a smart city development project was to reduce environmental impact to a sustainable level within the traditional urban infrastructure setting (Huang et al., 2021). Finally, socioeconomic factors tend to drive residents towards a smart city configuration, anticipating higher salaries, lower commute costs, and superior service delivery (Nathaniel, 2021). However, subsidies and incentives that finance smart city initiatives may also challenge socioeconomic equity distribution among city inhabitants, strengthening its relevance (Goel & Vishnoi, 2022).

**DATA**

The study integrated quantitative and qualitative data to understand the subject matter comprehensively.

Quantitative primary data derived through the survey included responses from a diverse demographic of city residents. Of the (n=50) sample, thirty respondents were from New York City, and twenty were from Seattle.

Qualitative data was sourced from multiple reputable journal platforms through the New York University Library, which included ScienceDirect, Jstor, SpringerLink, and MDPI.

| Variable | Variable Name | Journal Name | Author | Source |
|---|---|---|---|---|
| Y | Urbanization Patterns | *SpringerLink* | (Kundu & Pandey, 2020) | Secondary |
| Y | Urbanization Patterns | *ScienceDirect* | (Bounoua et al., 2018) | Secondary |
| Y | Urbanization Patterns | *IEEE Xplore* | (M. Shahidehpour et al., 2018) | Secondary |
| $X_1$ | Smart Technology | *ScienceDirect* | (Kummitha, 2018) | Secondary |
| $X_1$ | Smart Technology | *Central and Eastern European Library* | (Popescu, 2015) | Secondary |
| $X_1$ | Smart Technology | *IOP Science* | (Okafor et al., 2021) | Secondary |
| $X_2$ | Government Policy | *ScienceDirect* | (Mohanty & Kumar, 2021a) | Secondary |
| $X_2$ | Government Policy | *SpringerLink* | (Ahani & Dadashpoor, 2021) | Secondary |
| $X_2$ | Government Policy | *ScienceDirect* | (Angelidou, 2014) | Secondary |
| $X_3$ | Environmental Sustainability | *Wiley* | (Morrison et al., 2017) | Secondary |
| $X_3$ | Environmental Sustainability | *SpringerLink* | (Chivas et al., 2023) | Secondary |
| $X_3$ | Environmental Sustainability | *ScienceDirect* | (Shamsuzzoha et al., 2021) | Secondary |
| $X_4$ | Socioeconomic Factors | *Taylor&Francis* | (Angelo & Vormann, 2018) | Secondary |
| $X_4$ | Socioeconomic Factors | *IEEE Xplore* | (T. Persaud et al., 2020) | Secondary |
| $X_4$ | Socioeconomic Factors | *ScienceDirect* | (Esposito et al., 2021) | Secondary |

*Table 1: Qualitative data sources for each of the study's variables*

Table 1 contains a list of journals referenced during research and its correlation to the study's variables.

**FRAMEWORK**

The Theory of Urban Complexity, pioneered by Jane Jacobs in 1961, suggests that cities are dynamic, complex systems that interact with various elements and are always self-organizing (Jacobs, 1961). Complexity, Successive Assumption, and Simplicity are the particularly influential theory concepts, according to Jacobs (Jacobs, 1961). The theory posits that urban systems are not linear but evolve through the interactions between these elements, leading to emergent behaviors and patterns. It emphasizes the importance of considering urban systems' interdependencies and feedback loops, which are fundamental to informing the research study's dependent Y variable, urbanization patterns (Perrone, 2019).

The theory explains the relationships with smart city technology ($X_1$) and how it contributes to urban systems' emergent behaviors, such as changes in mobility patterns, energy consumption, or civic engagement. It suggests that technologies act as enablers and disruptors, potentially leading to new urban forms and functions (Mohanty & Kumar, 2021a).

Concerning government policies ($X_2$), the theory explains how policies designed to promote smart city initiatives can influence urbanization by facilitating or constraining the integration of technologies (Sengupta & Sengupta, 2017). Policies that support integrating smart technologies, protect diversity, and encourage mixed-use developments are crucial. Government policies can either foster the organic complexity proposed by Jacobs (1961) by enabling success factors or stifle it through rigid, prescriptive regulations. Jacobs argued for policies that respect the organic, bottom-up processes that contribute to urban vitality and guide development toward sustainability and inclusivity (Sengupta & Sengupta, 2017).

Jacobs's (1961) initial theory implicitly supported environmentally sustainable ($X_3$) urban development, emphasizing compact, mixed-use, pedestrian-friendly neighborhoods. Other scholars have since expanded this assertion to address evolving, contemporary environmental concerns (Ekka et al., 2024). By highlighting the role of environmental sustainability in maintaining the resilience and health of urban systems, the theory provides a comparative perspective between smart city initiatives and ecological outcomes (Jacobs, 1961).

The theory of urban complexity also addresses socioeconomic factors ($X_4$) by explaining how disparities in access to technology, economic opportunities, and social equity influence urbanization patterns, revealing the importance of designing smart cities that are equitable and inclusive (Wright, 2015).

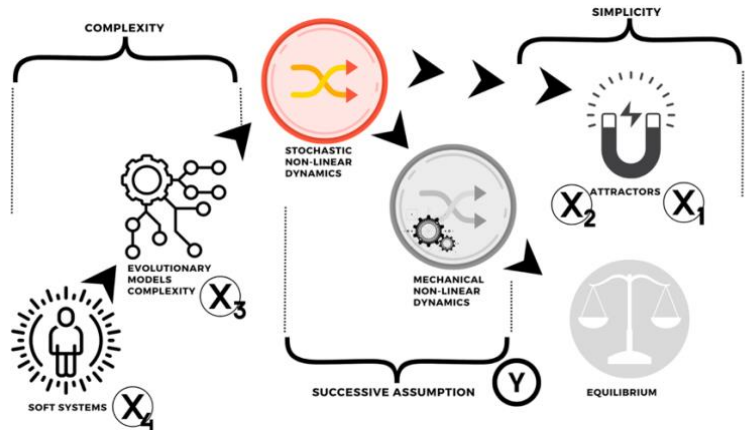

*Figure 3: Urban complexity theory vs smart city impact on urbanization. Data Source: Portugali et al. (2012)*

The theory of urban complexity, depicted in Figure 3, posits that the components of an urban population are self-regulating, attending to deficiencies while creating new ones (Portugali et al., 2012). The theory comprises three phases: Complexity, Successive Assumption, and Simplicity.

| Theory Category | Theory Component | Variable |
|---|---|---|
| Complexity | Soft Systems | $X_4$ - Socioeconomic |
| Complexity | Evolutionary Models Complexity | $X_3$ – Environmental Sustainability |
| Successive Assumption | Stochastic Non-linear Dynamics | Y- Urbanization Patterns |
| Successive Assumption | Mechanical Non-linear Dynamics | Y- Urbanization Patterns |
| Simplicity | Attractors | $X_1$ – Smart City Technology $X_2$ – Government Policy |
| Simplicity | Equilibrium | $X_{1234}$ and Y |

*Table 2: Urban complexity theory components*

Table 2 illustrates the categories and components of the urban complexity theory and its relevance to the research study. Each of the theory's categories and related components is correlated to the study's variables, as indicated, to appropriately inform the research (e.g., Complexity with Environmental Sustainability and Successive Assumption with Urbanization Patterns).

Soft systems describe elements that inform evolutionary models. These systems are not scientific but are supported by literature, institutions, or history (Portugali et al., 2012). Independent variable $X_4$, socioeconomic factors, represents a soft system reality as they are primarily influenced by historical events, as described by Portugali et al. (2012).

Most complexity theorists, including Jacobs (1961), categorize evolutionary models as catalysts by, often referred to as "Emergence." The process is the interaction among smaller entities or innovations to produce a new noticeable entity or subsequent dynamic model (Sengupta & Sengupta, 2017). An example is the clustering of high-rise residential buildings near an arterial train station, which leads to a spike in carbon emissions from the increased business concentration. In this case, environmental sustainability (independent variable $X_3$) is the resultant evolutionary model variable contributing to change (Sengupta & Sengupta, 2017).

The Successive Assumption category is representative of the study's dependent variable, urbanization patterns (Y). Stochastic models align with the theory by capturing urban systems' unpredictable, emergent nature where small, random events can disproportionately affect urbanization patterns (Jacobs, 1961). These models can be referenced to help explain how unpredictable factors contribute to the diversity and resilience of cities, influencing where people live and work and how urban spaces evolve. However, mechanical non-linear dynamics models reflect urban systems' structured but susceptible and interconnected nature. They illustrate how specific urban policies, infrastructure developments, or socioeconomic changes may lead to non-linear responses in urbanization patterns, such as rapid urban sprawl or concentrated urban renewal, highlighting the critical thresholds and feedback loops that drive urban change (Portugali et al., 2012).

The Simplicity component consists of attractors and stationary equilibrium. Attractors are sets of states (or outcomes) a system draws towards over time (Portugali et al., 2012). In urban systems, attractors can represent stable patterns of urban development, lifestyle choices, or economic activities that emerge from the interactions of various urban elements. Attractors can be points (single states), cycles (repeating patterns), or even strange attractors (complex patterns), depending on the dynamics of the system. Stationary Equilibrium refers to a state of balance where the system's overall state remains constant over time despite ongoing processes and interactions. Urban complexity theory could imply a stable population distribution, a steady-state economy, or a balance between urban expansion and green space preservation (Portugali et al., 2012).

Smart city technologies (independent variable $X_1$) can act as modifiers or disruptors of existing attractors in urban systems. For instance, implementing intelligent transportation systems could shift urban attractors related to mobility, encouraging more sustainable transport behaviors and potentially leading to new equilibrium states characterized by reduced traffic congestion and pollution (Portugali et al., 2012). Similarly, smart infrastructure can facilitate the emergence of smart grids and sustainable energy usage patterns, altering attractors related to energy consumption and leading urban systems toward new, sustainable stationary equilibriums (Sengupta & Sengupta, 2017).

Government policies (independent variable $X_2$) can deliberately shape urban attractors by incentivizing certain behaviors or developments and discouraging others. Zoning laws, urban planning regulations, and policies promoting smart city growth are examples of how government actions can modify the landscape of urban attractors (Sengupta & Sengupta, 2017). For instance, policies encouraging mixed-use development can lead to attractors favoring vibrant, diverse urban spaces, moving the city towards a stationary equilibrium that balances residential, commercial, and recreational uses. Government policies can also establish or shift attractors related to social equity and accessibility, influencing how technology benefits are distributed across the urban population (Sengupta & Sengupta, 2017).

## SAMPLE

Two of the most technologically mature cities in the U.S. were selected to assess smart cities' impact on urbanization: New York City and Seattle (Shah et al., 2019). The researcher canvased professional organizations to distribute the survey, which included city planners, analysts, developers, and residents of densely populated suburbs. Manhattan and Queens were selected for New York, and Capitol Hill was chosen for Seattle. The following zip codes were included in the distribution sample: 10001, 10009, 11104, 11105, 98102, 98112, and 98122.

The target sample was (n=50), of which 53% were males and 47% females. The total sample allocation represents a 60/40 split between New York and Seattle, respectively, with the survey allotted evenly within each city's suburbs. Twenty-five percent of the survey comprised demographic questions to assess sample diversity.

None of the respondents were under 18 or had diminished autonomy. The target response age was 18 to 65.

**INSTRUMENT**

A web-based experiment was conducted where 50 participants from New York City and Seattle were presented, following their consent to participate in the study, with a survey to express their smart city opinion. The survey included five demographic questions to record their age, sex, education level, employment status, and suburb of residence. The survey also included fifteen questions adapted from Molná's (2021) methodology practice on smart city experience questionnaires. All non-demographic questions followed a 5-point Likert scale ranging from strongly agree to strongly disagree (i.e., strongly agree, somewhat agree, neutral, somewhat disagree, and strongly disagree). These questions assessed participants' perceptions of smart city projects' effectiveness, challenges, and benefits in influencing urban development patterns.

Qualtrics, a secure online survey tool, was selected to aggregate the data from the responses and analyze the effects of the independent variables on the dependent variable. The survey followed a closed-type format with access by electronic invitation only. It was preceded by an informed consent and disclaimer clearly stating that the purpose of the study was for student research only and that the data and identity of the participant would be kept confidential.

**SURVEY**

The survey was developed by integrating key aspects from Molná's (2021) methodology practice on smart city experience questionnaires, focusing on understanding perceptions towards smart city technologies ($X_1$), government policies ($X_2$), environmental sustainability ($X_3$), and socioeconomic factors ($X_4$) in urbanization patterns. The questions were deliberately structured to transition from specific independent variables (smart city technology, government policy, environmental sustainability, socioeconomic factors) toward a broader understanding of general perceptions and demographic backgrounds. This progression aimed to gather targeted insights on each independent variable affecting urbanization patterns and contextualize these insights into the wider narrative of urban development and personal demographics.

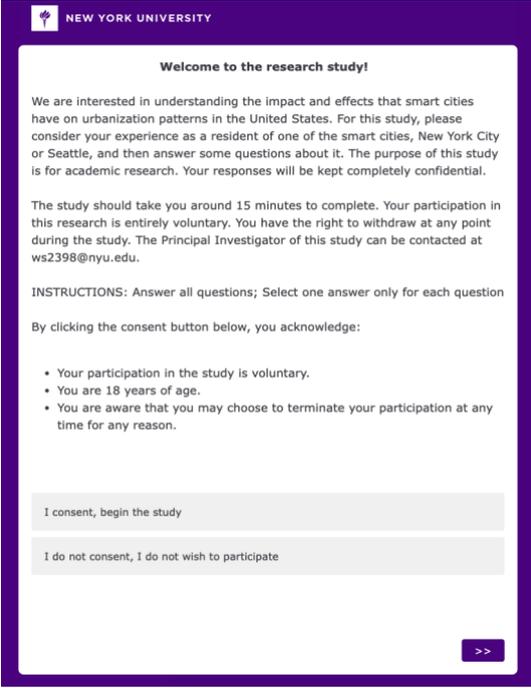

*Figure 4: Informed Consent Acknowledgment*

As indicated in Figure 4, the informed consent form was a preamble to the survey, which informed the participant about the study's directive and provided instructions on completing it.

| Question | Variable |
|---|---|
| Question 1 | Y |
| Question 2 | $X_1$ |
| Question 3 | $X_1$ |
| Question 4 | $X_2$ |
| Question 5 | $X_2$ |
| Question 6 | $X_3$ |
| Question 7 | $X_3$ |
| Question 8 | $X_1$ |
| Question 9 | $X_4$ |
| Question 10 | $X_4$ |
| Question 11 | $X_4$ |
| Question 12 | $X_2$ |
| Question 13 | $X_4$ |
| Question 14 | $X_4$ |
| Question 15 | $X_3$ |

*Table 3: Question/Variable Key*

Table 3 describes the key to which variable was assessed for each descriptive, non-demographic survey question, as indicated in Figure 5.

| Q# | Question |
|---|---|
| Q1 | The population in my area has noticeably increased in the last two years |
| Q2 | I believe smart city technologies contribute to making urban living more sustainable. |
| Q3 | Smart city solutions have enhanced public safety and security in my area. |
| Q4 | City initiatives and policies are clear and transparent in driving smart city developments in my area. |
| Q5 | I feel adequately informed and engaged in the planning and execution of smart city projects by the local government |
| Q6 | Smart city projects in my city have positively impacted the local environment and green spaces. |
| Q7 | Parks and green spaces are abundantly accessible in my city |
| Q8 | Public transport is always operating and available when required in my city |
| Q9 | Income controlled housing is accessible through most establishments in my city |
| Q10 | There is a noticeable homeless population across my city |
| Q11 | I get paid a higher salary for my job in my city than if I worked that same job elsewhere in the US |
| Q12 | My city is continually developing new services in my area to meets the community's needs and expectation |
| Q13 | Smart city services have improved my work or personal life |
| Q14 | The cost of living in my city is significantly more expensive than anywhere else in the US |
| Q15 | I do not intend to move from my city within the next five years |

For Q1-Q15, response options: Strongly Agree, Somewhat Agree, Neutral, Somewhat disagree, Strongly disagree.

Q16 — What is your age: 18-30, 31-40, 41-50, 51-60, >60

Q17 — What is your gender: Female, Male

Q18 — What is your education level: Less than High School, High School, Some college, Associate degree, Bachelors degree, Masters degree, Doctorate degree

Q19 — What is your employment status: Full-time employed, Part-time employed, Retired, Seeking Work, Student, Unable to work

Q20 — What is your city of residence: Manhattan, New York; Capitol Hill, Seattle

*Figure 5: Survey Questionnaire – Smart City Impact on Urbanization Patterns in the US*

The questions were drafted using NYU's Qualtrics survey system and included native statistical tools to interrogate the response. As indicated in Figure 5, questions 1 through 15 are descriptive and integrate assessment of the independent and dependent variables. Questions 16 to 20 are demographic.

## FIELDWORK

The survey was a closed, cross-sectional study conducted electronically. Access to the Qualtrics survey was managed by communicating a secure link to the participant's disclosed email address. Although the researcher was responsible for managing the survey distribution to participants, assistance was provided by organizations in New York City and Seattle. These include New York and Seattle city planning departments, Samsara, Cavnue, Manhattan Community Board, and North Capitol Hill Organization. A representative of each organization distributed the consent and questionnaire to each approved participant via their email address.

## STATISTICAL TOOLS

This study's analysis of data collected from surveys and secondary sources employs various statistical tools to interpret quantitative and qualitative data to understand how smart cities effectively influence urbanization patterns. Descriptive statistics methods summarize the demographics of the study's participants and their responses to survey questions. These statistics provide an overview of the sample characteristics and the distribution of responses across the Likert scale questions. Central midpoint measurement markers such as mean, median, and mode were used to identify common trends or patterns in responses, offering initial insights into the attitudes and experiences of the sample.

Regression analysis aided in evaluating the relationship between the independent and dependent variables. The method included four steps: the initial step discussed the strength of the relationship between the dependent and independent variables, indicating the proportion of variance in urbanization patterns explained by the smart city initiatives; the second step evaluated the overall fit of the regression model, determining whether the model is statistically significant in explaining changes in urbanization patterns; in the third step, the analysis focused on the impact (positive or negative) of each independent variable on urbanization, based on the regression coefficients; and the final step assessed the statistical significance of each independent variable's impact on urbanization, with variables considered significant if the p-value is less than or equal to 0.05.

Integrating descriptive statistics and regression analysis findings seeks to validate the dynamics between smart cities and urbanization. By employing these statistical tools, the research highlights significant trends and patterns and explores the underlying smart city mechanisms driving urbanization.

This comprehensive statistical approach ensures the study's conclusions are robust, evidence-based, and actionable, contributing valuable insights to the ongoing discourse on smart cities and urban development.

### Definition of Terms

**Attractors:** represent specific sets of states toward which a system tends to evolve, regardless of its starting conditions. Depending on the system's nature, attractors can be points, curves, manifolds, or even more complex structures known as strange attractors. The key characteristic of an attractor is its ability to "attract" trajectories in its phase space (the space of all possible states of the system) over time, leading to predictable patterns of behavior within the system (Portugali et al., 2012).

**Complexity:** source of creative interaction innovation and change concerning urbanization (Portugali et al., 2012).

**Environmental Sustainability:** the responsible interaction with the planet to maintain natural resources and avoid jeopardizing the ability of future generations to meet their needs. It involves making decisions and taking action that are in the interest of protecting the natural world, with particular emphasis on preserving the capability of the environment to support life. (World Commission, 2014).

**Equilibrium:** is the state in which a system is in balance and experiences no net change over time. Depending on the system's nature and the forces acting upon it, equilibrium can manifest in various forms (Gonzalez, 2017).

**Government Policy:** refers to the plans and actions undertaken by a government to achieve specific societal goals. These policies can encompass various areas, including economic, social, environmental, and administrative domains. Government policies are formulated through legislation, regulations, decisions, and actions of institutional bodies and officials and are implemented to guide the behavior of citizens, businesses, and other government agencies. Policies are designed to address public issues, manage resources, protect citizens, and promote societal welfare and development (Rochefort, 1997).

**Mechanical Non-linear Dynamics:** refers to the study and analysis of systems whose behavior cannot be accurately described by linear equations due to non-linear relationships among system components. These dynamics are deterministic, meaning that given a specific set of initial conditions, the system's future behavior can be predicted, albeit the relationships and outcomes are complex and may exhibit sensitivity to initial conditions, leading to diverse behaviors even from small changes at the start (Strogatz, 2018).

**Socioeconomic Factors:** the social and economic characteristics that define society's individual and collective living conditions. These factors encompass various attributes related to an individual's or population's financial status, education level, employment, income, wealth, and social standing (Leventhal & Brooks-Gunn, 2000).

**Soft Systems:** components of urban pattern influence not derived from science. This includes heuristics, intuition, literature, and historical events (Checkland & Poulter, 2007).

**Successive Assumption**: the iterative process of updating and refining models or theories based on new information and observations. As researchers and planners collect more data and gain deeper insights into how cities operate, they adjust their assumptions and models to better reflect urban dynamics' realities (Portugali et al., 2012).

**Stochastic Non-linear Dynamics:** the mathematical and conceptual approach to understanding systems that exhibit randomness (stochasticity) and non-linearity in their behavior and evolution. This approach models and analyzes systems where outcomes are not deterministic but are influenced by inherent unpredictability and complex interactions among multiple variables or components (Gardiner, 2009).

**Smart City:** an urban area that uses electronic Internet of Things (IoT) sensors to collect data, which is then used to manage assets and resources. This includes data collected from citizens, structures and infrastructure that are processed and analyzed to monitor and manage public services and entities (Neirotti et al., 2014).

**Smart City Technology:** refers to integrating and applying Information and Communication Technologies (ICT) and Internet of Things (IoT) devices to manage and optimize city services, infrastructure, and operations. These technologies aim to enhance the quality of life for residents, improve sustainability, increase efficiency in urban services, and facilitate more informed decision-making by city administrators and planners (Batty et al., 2012).

**Urbanization Patterns:** the specific ways urban growth and development unfold over time and space within cities and metropolitan areas. These patterns encapsulate the changes in land use, population distribution, economic activities, and the built environment resulting from urbanization (Seto et al., 2011)

## LITERATURE REVIEW

Existing qualitative data on urbanization trends in the United States informs a foundational understanding of the effects of rapid population density increases. Smart cities have emerged and developed in the most populous urban cities in response to increased demands on infrastructure, the environment, and service delivery of residents (DeHart et al., 2021).

| Variable | Variable Name | Journal Name | Author | Source |
|---|---|---|---|---|
| Y | Urbanization Patterns | *SpringerLink* | (Kundu & Pandey, 2020) | Secondary |
| Y | Urbanization Patterns | *ScienceDirect* | (Bounoua et al., 2018) | Secondary |
| Y | Urbanization Patterns | *IEEE Xplore* | (M. Shahidehpour et al., 2018) | Secondary |
| $X_1$ | Smart Technology | *ScienceDirect* | (Kummitha, 2018) | Secondary |
| $X_1$ | Smart Technology | *Central and Eastern European Library* | (Popescu, 2015) | Secondary |
| $X_1$ | Smart Technology | *IOP Science* | (Okafor et al., 2021) | Secondary |
| $X_2$ | Government Policy | *ScienceDirect* | (Mohanty & Kumar, 2021a) | Secondary |
| $X_2$ | Government Policy | *SpringerLink* | (Ahani & Dadashpoor, 2021) | Secondary |
| $X_2$ | Government Policy | *ScienceDirect* | (Angelidou, 2014) | Secondary |
| $X_3$ | Environmental Sustainability | *Wiley* | (Morrison et al., 2017) | Secondary |
| $X_3$ | Environmental Sustainability | *SpringerLink* | (Chivas et al., 2023) | Secondary |
| $X_3$ | Environmental Sustainability | *ScienceDirect* | (Shamsuzzoha et al., 2021) | Secondary |
| $X_4$ | Socioeconomic Factors | *Taylor&Francis* | (Angelo & Vormann, 2018) | Secondary |
| $X_4$ | Socioeconomic Factors | *IEEE Xplore* | (T. Persaud et al., 2020) | Secondary |
| $X_4$ | Socioeconomic Factors | *ScienceDirect* | (Esposito et al., 2021) | Secondary |

*Table 3: Literature review sources*

The following literature review derived from existing studies in Table 3 confirms that smart city solutions positively impact urbanization problems in the United States. Therefore, specific smart city programs and incentives should be established to support the exponential growth of urbanization within U.S. cities.

Urbanization Patterns in the United States – Y Variable (dependent variable)

*How do smart cities impact urbanization patterns(Y) in the United States?*

The academic journals selected for the dependent variable review, as in Table 3, were sourced from credible peer-reviewed platforms to introduce qualitative data that informs fundamental categories of knowledge on urbanization trends. For example, benchmarking the world trend in population movement of an urban configuration is vital in performing a comparative analysis of patterns in the United States. In addition, analyzing the physical changes of a dense urban smart city landscape offers valuable insight into environmental and socioeconomic impacts.

Bounoua et al. (2018) identified common characteristic changes in the geographic landscape of a city as urbanization increased by eleven percent in the last decade. The most apparent of these characteristics is the positive relationship between a city's population growth and the expansion of vertical structures. Bounoua et al. (2018) confirm that high-rise modern buildings displace traditional structures as population density increases – a lever in shifting to a smart city configuration. This shift to urban living for the world's population is occurring rapidly, as confirmed by Kundu and Pandey (2020), with the United States leading the trend for the past seventy years and projected to do so for the next twenty-five years (Bounoua et al., 2018).

Although studies from Bounoua et al. (2018), Kundu and Pandey (2020) support the findings that indicate urbanization patterns are representative of changes in population distribution, urban area expansion, and land use and infrastructure transformation, the impact of the densest configuration, a smart city, is relatively unknown. For example, M. Shahidehpour et al. (2018) concluded that human-machine partnerships in a smart city setting had notable benefits. However, the conclusion did include significant caveats. The infancy of the smart city concept and the cybersecurity risks associated with technology were among the study's concerns (M. Shahidehpour et al., 2018). Technology advancement, however, from 2018 to 2023 occurred at a rate of more than one hundred and four percent – a significant factor in smart city scaling, which M. Shahidehpour et al. (2018) did not account for (*Global IT Industry Growth Rate by Segment 2018-2023*, n.d.). Addressing unknowns within the study concludes which smart city components promote and support urbanization and subsequently address the problem of sustaining growing populations within dense urban cities in the United States (M. Shahidehpour et al., 2018).

<u>Smart City Technology – $X_1$ (independent Variable)</u>

*Does integrating smart city technology in urban planning ($X_1$) impact urbanization patterns in the United States (Y)?*

      Smart city technology, characterized by a broad suite of digital innovations including IoT devices, advanced data analytics, blockchain, and Artificial Intelligence (AI), plays a pivotal role in optimizing city functions, enhancing public services, enabling sustainable development, and stimulating economic growth (Popescu, 2015).

      Kummitha (2018) analyzed the role of technology in successful smart city innovation projects and found it to be one of the critical factors in fostering urban transformation within the United States. Kummitha's (2018) study concluded a positive correlation between the integrated implementation of smart technology infrastructure and the city's success at addressing contemporary challenges such as urban congestion, energy consumption, and public service delivery. Therefore, Kummitha's

(2018) findings confirm that smart technology contributed to increased urbanization concentration.

Popescu (2015) further complements this conclusion by exploring the economic dimensions of smart city technologies. The emphasis on how these innovations catalyze economic development and innovation ecosystems elucidates their indirect influence on urbanization. For example, Popescu (2015) found that enhanced connectivity and advanced ICT infrastructure attract businesses and talent, fostering an environment that encourages innovation and job creation. Consequently, these economic incentives drive population movements towards technologically advanced urban centers, shaping urbanization trends (Popescu, 2015).

The United Nations estimates the urbanized world population will grow from fifty-five percent to sixty-five percent by 2025; it is, therefore, valuable to identify the technology that is instrumental in facilitating the anticipated growth (Okafor et al., 2021). Okafor et al. (2021) concluded through a meta-analysis that eight specific technologies are critical to support the maturity of smart cities and scaled urbanization: transportable smart city infrastructure, financial technology, self-repairable smart city components, online education, digital government treasury services, driverless cars, self-detecting pandemic technology, and national electronic voting systems (Okafor et al., 2021).

It is evident from Popescu (2015), Kummitha (2018), and Okafor et al. (2021) that smart city technology significantly influences urbanization trends and migration patterns due to its integration into urban planning and economic development.

Government Policy – $X_2$ (Independent Variable)

*Does government policy ($X_2$) influence urbanization patterns (Y) in the context of smart cities?*

Policy and governance frameworks are the rules and actions by which urbanization, including smart city initiatives, are guided and implemented (Schlager, 2019). Government policies, encompassing regulatory frameworks, funding mechanisms, strategic planning, and public-private partnerships, are pivotal in steering the development and integration of smart technologies into urban environments (Mohanty & Kumar, 2021).

Mohanty & Kumar (2021) identified a high success rate between targeted government interventions, such as fiscal incentives, policy support, infrastructure investment, and smart city projects. Mohanty & Kumar (2021) articulate that these policy interventions foster innovation and technological deployment and positively influence urbanization patterns by incentivizing sustainable business development, enhancing urban resilience, and improving public services.

Conversely, Ahani & Dadashpoor (2021) studied the efficacy of growth containment policies as an attempt by governments to slow rapid urbanization movement, especially in peri-urban areas (rural to urban transition locations). Ahani & Dadashpoor (2021) concluded that instruments such as taxes, limitations on financial rights, and property ownership incentives in rural towns were ineffective in slowing urbanization growth. The study confirmed that urban containment policy has a negligible effect on slowing urbanization growth (Ahani & Dadashpoor, 2021).

Angelidou (2014) established that one of the most important policy implications for smart city urbanization is spatial reference. Spatial reference is a smart city's geographical landscape and physical features that influence four key policy relationships: national versus local strategies, new versus existing cities, hard versus soft infrastructure-oriented strategies, and economic sector-based versus geographically-based strategies (Angelidou, 2014). Although Angelidou (2014) identified that each strategic direction has benefits and disadvantages, every smart city strategy must be supported by policies promoting digital changes. Furthermore, they should consider institutional variances to achieve a balance between technological advancements and social cohesion (Angelidou, 2014).

      Favorable government policy is an important influencing factor for promoting the resilience of a smart city (Mohanty & Kumar, 2021a). However, Ahani & Dadashpoor (2021) concluded that policies with an urban containment directive are counterproductive and encourage urbanization movement to a smart city. Angelidou (2014) found that irrespective of the policy adopted for new or existing smart cities, adaptation to spatial reference and institutional fluctuations are critical factors in sustainable urban planning. Government policy, therefore, is significant in influencing urbanization trends.

Environmental Sustainability – $X_3$ (Independent Variable)

*Does environmental sustainability ($X_3$) impact urbanization patterns (Y) in the context of smart cities?*

      The environmental aspect in the context of smart cities includes sustainability practices and the impact of urban development on natural ecological resources. Analyzing how smart city designs and technologies aim to reduce environmental footprints and assessing their practicality and sustainability is essential in understanding the impact on urbanization trends.

Morrison et al. (2017) concluded that integrating environmental sustainability into smart city initiatives is essential to the health and well-being of urban ecosystems and their inhabitants as urban areas expand. Morrison et al.'s (2017) findings also point to problems of excessive cumulative carbon emission levels and a high water contamination rate as byproducts of densely populated smart cities. According to Morris et al. (2017), these ecological challenges are effectively addressed when incorporating ecosystem services into a smart city's initial urban planning and development. By integrating natural elements and systems into the smart city strategy, Morris et al. (2017) concluded that cities could mitigate adverse environmental impacts and improve air and water quality.

Chivas et al. (2023) furthered this discourse by presenting a comprehensive framework for integrating technology and ecology within smart cities. According to Chivas et al. (2023), the pace and complexity of population and business growth of a smart city create an ecological impact challenge that renders traditional approaches ineffective. An alternative solution, as posited by Chivas et al. (2023), is a system called "Green Mesh." The system monitors suburbs and blocks within the city to detect abnormally high carbon emissions. It then recommends planting specific plants or vegetation within the affected area to stabilize gas emissions (Chivas et al., 2023). Chivas et al. (2023) confirm that the "Green Mesh" solution, which integrates into existing smart city infrastructure using Industrial Internet of Things (IIoT) sensors, has effectively limited or reduced the carbon emission growth in residential areas of a smart city (Chivas et al., 2023).

Shamsuzzoha et al. (2021) examined three leading smart cities worldwide, London, Helsinki, and Singapore, and concluded that all three locations met or exceeded environmental guidelines set forth by the International Organization for Standardization (ISO) for urban environment sustainability. Shamsuzzoha et al. (2021) evaluated seventeen categories, including energy, health, wastewater, transportation, solid waste, and safety. Shamsuzzoha et al. (2021) concluded that smart cities significantly outperformed traditional urban living, which contributed to consistently sustainable living conditions.

Environmental sustainability is central to the development of any smart city. The dense concentration of people and the businesses supporting their needs generate substantial ecological waste. As confirmed by Morris et al. (2017), environmental impact informs the decisions by city planners when allocating business and residential zones, which subsequently influence the urbanization decisions of individuals. Continuous improvement or sustainability of a city's ecological environment is attractive to potential residents, as concluded by Chivas et al. (2023) and Shamsuzzoha et al. (2021), therefore influencing urbanization patterns in the context of a smart city.

<u>Socioeconomic Factors – $X_4$ (Independent Variable)</u>

*Do socioeconomic ($X_4$) factors impact urbanization patterns (Y) of smart cities?*

Socioeconomic dynamics, encompassing income levels, education, employment, and social inequality, play a pivotal role in shaping the development and success of smart city initiatives.

The progress toward the formation of a smart city was catalyzed by the historical evolution of urbanization reforms, according to Angelo and Vormann (2018). Furthermore, Angelo and Vormann's (2018) findings correlate socioeconomic aspirations and challenges to six cycles of urban reform. For example, the third cycle that gave rise to industrialization between 1875 and 1918 led to stark disparities in social inequality (Angelo & Vormann, 2018). Angelo and Vormann (2018) found this resulted in a "city efficient movement" where the government's focus was technology, governance, and economic growth to improve the quality of urban city living. In the sixth cycle from 2012 to now, Angelo and Vormann (2018) concluded that globalization was the most significant contributor to modern socioeconomic disparity. The smart city was, therefore, a response to the "crisis" of the fifth cycle, which included the offshoring of manufacturing jobs and multiple other global problems that prioritized efficiency to achieve liveable cities during an economic downturn (Angelo & Vormann, 2018).

Persaud et al. (2020) concluded that individuals determine the quality of life in an urban setting based on transportation, social networks, safety and mobility, and healthcare. Furthermore, in an urban smart city environment, there is a focus on measuring the efficacy of gamification, according to Persaud et al. (2020), which measures how well quality of life metrics are integrated into a smart city. Persaud et al.'s (2020) study further compared urbanization between smart city countries and those of traditional towns and concluded that the migration rate to smart cities was more significant when compared to conventional cities.

Esposito et al. (2021) found that residents of the smart city of Brussels prioritized three socioeconomic factors when considering where to live: sustainable development, democratic governance, and security. Esposito et al. (2021) concluded that these driving factors influenced migration from rural and traditional non-smart neighboring towns to Brussels.

Socioeconomic factors, therefore, directly affect the urbanization patterns of smart cities. Persaud et al. (2018) concluded that entrepreneurial opportunities and better job prospects drive urbanization due to perceived better quality of life within a smart city, and Esposito et al. (2021) further support the inference by concluding that residents seeking urban security and democratic governance prefer smart cities when compared to traditional alternatives. Conversely, as Angelo and Vormann (2018) concluded, economic downturns and social inequality attract people to the efficiency of a smart city.

## Data Analysis

Sociodemographic Profile of Sample

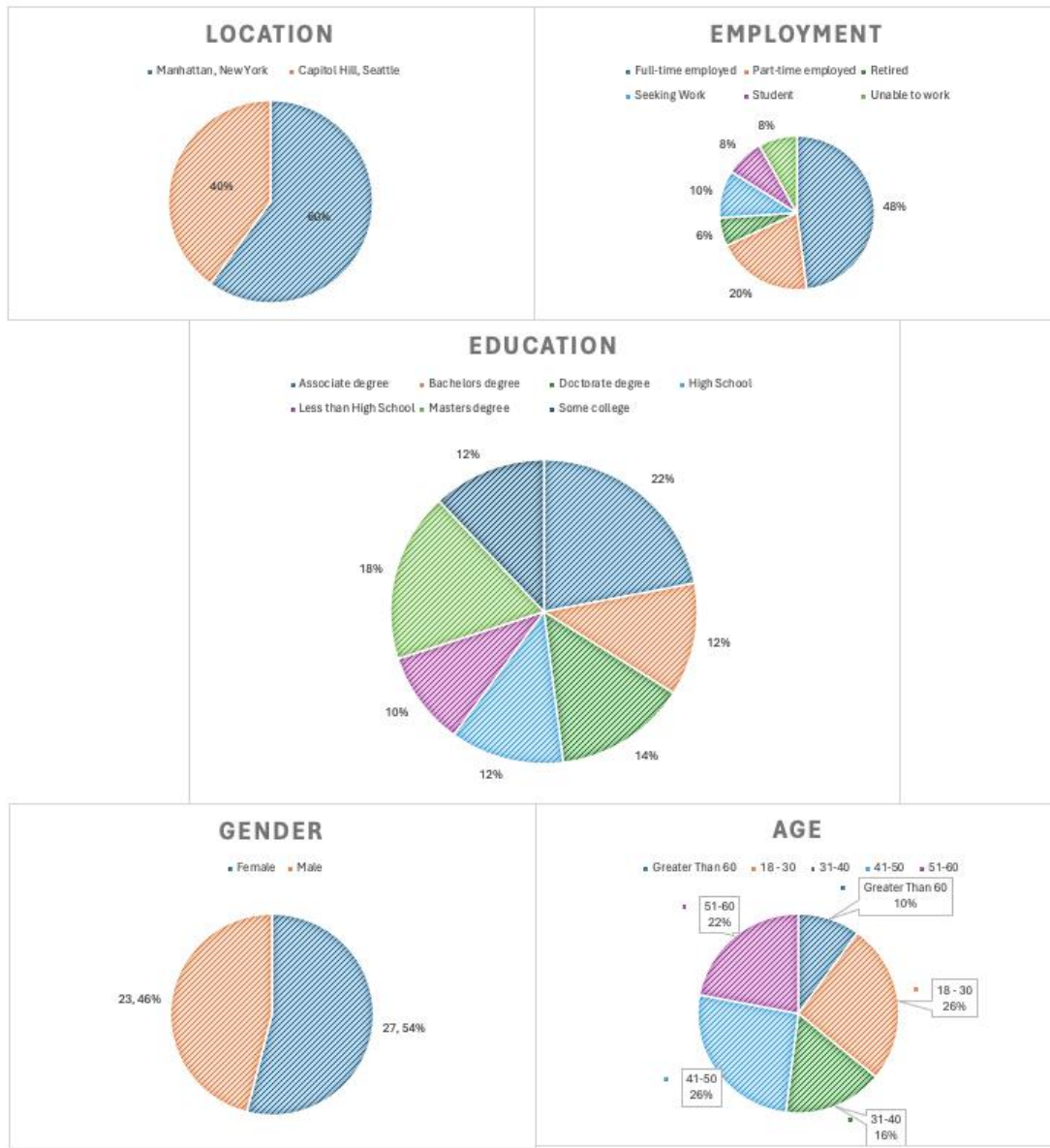

*Figure 6: Pie charts illustrating sociodemographic variables: employment, education, age, gender, location*

The sociodemographic profile of the sample is summarized in Figure 6. It highlights a varied yet notably middle-aged, educationally diversified, and geographically distinct group predominantly employed full-time. Gender representation was closely matched, with males accounting for 54% and females at 46%. Age distribution indicates a spread across various adult life stages, with 26% of participants aged 18-30 and 26% within the 41-50 range, suggesting a sample well distributed across working-age adults. The participants aged 51-60 constitute a significant 22%, while those over 60 represent a smaller segment at 10%.

The sample's geographical target was split between Manhattan, New York, 60%, and Capitol Hill, Seattle, 40%, providing a significant urban influence within the responses.

Education levels within the sample are widely ranged, with the most common being associate degrees and some college, each making up 22% of the sample. Participants with bachelor's degrees represent 18% and master's degree holders 14% of the sample. Those with less than high school education and doctoral degrees form an equally small proportion, comprising 12%, with the least represented being high school graduates at 10%.

Employment status is predominantly full-time at 48%, suggesting economic stability, with part-time employment at 20%. Retirees comprise 10% of the sample, reflecting the post-employment population. Additionally, students form 8% of the participants, providing insight into the younger, potentially academically engaged demographic. Those actively seeking and unable to work comprise 6% of the sample, highlighting the variance in employment status within the group.

The sample portrays a snapshot of an adult demographic with a balanced gender distribution, a concentration in significant urban centers, and various educational backgrounds, mainly within the employed sector. The detailed breakdown of these sociodemographic characteristics can provide valuable context for understanding the survey responses and informing conclusions.

## DESCRIPTIVE STATISTICS RESULTS

| SURVEY NUMBER | Q1 | Q2 | Q3 | Q4 | Q5 | Q6 | Q7 | Q8 | Q9 | Q10 | Q11 | Q12 | Q13 | Q14 | Q15 |
|---|---|---|---|---|---|---|---|---|---|---|---|---|---|---|---|
| VARIABLE | Y | X1.Smart Technology | X1.Smart Technology | X2.Government Policy | X2.Government Policy | X3. Environment | X3. Environment | X1.Smart Technology | X4.Socio economic | X4.Socio economic | X4.Socio economic | X2.Government Policy | X4.Socio economic | X4.Socio economic | X3. Environment |
| MEAN | 4 | 4 | 4 | 3 | 4 | 4 | 3 | 3 | 3 | 3 | 3 | 4 | 4 | 4 | 4 |
| MEDIAN | 4 | 4 | 4 | 4 | 4 | 4 | 3 | 3 | 3 | 3 | 4 | 4 | 3 | 4 | 4 |
| MODE | 5 | 5 | 4 | 5 | 4 | 5 | 1 | 3 | 3 | 3 | 4 | 4 | 5 | 4 | 4 |
| STANDARD DEVIATION | 1.400292 | 1.137308 | 1.140891166 | 1.356616439 | 1.161280082 | 1.25779204 | 1.454619664 | 1.336458922 | 1.547084157 | 1.300549 | 1.230397314 | 1.26571751 | 1.216216952 | 1.043737398 | 1.214873807 |

*Figure 7: Descriptive analytics of the sample*

Figure 7 describes the mean, mode, median, and standard deviation for each descriptive question of the survey. The calculated values can provide insight into the data's central tendency of the dependent and independent variables. For example, to determine whether a positive, neutral, or negative perception exists toward the variable assessed and associated with each question. The analysis of the results in Figure 7 is the source reference for the variable analysis below.

**Y Urbanization Patterns – Dependant Variable (Q1)**

*"Q1 -The population in my area has noticeably increased in the last two years."*

The mode of 5 indicates a strong positive sentiment that urbanization pattern changes are evident in the respondents. This is further reinforced by the median of 4, where half of the participants agreed with the sentiment. The mean of 4 also supports this conclusion. The standard deviation of 1.40029 indicates some variability in responses but not wide enough to discard consensus among respondents.

**X$_1$ Smart City Technology (Q2, Q3, Q8)**

*"Q2 – I believe smart city technologies contribute to making urban living more sustainable."*

*"Q3 – Smart city solutions have enhanced public safety and security in my area."*

The mean of 4 for both questions indicates a positive overall sentiment toward smart technology's impact. The median is also 4, which reinforces the positive sentiment. The standard deviation of approximately 1.2 for the question suggests a slight to moderate variation of opinions among respondents, though not highly divergent. Respondents likely see the benefits of smart technology within their city.

*"Q8 – Public transport is always operating and available when required in my city."*

The mean, mode, and median for this question are all 3, indicating a robust neutral opinion on transport among respondents. This may suggest that participants prefer not to answer the question, or another factor may influence their indifferent response. The standard deviation of 1.34 indicates opinions on both sides of the neutrality to agree or disagree.

**$X_2$ Government Policy (Q4, Q5, Q12)**

*"Q4 – City initiatives and policies are clear and transparent in driving smart city developments in my area."*

*"Q5 – I feel adequately informed and engaged in the planning and execution of smarty projects by local government."*

*"Q12 – My city is continually developing new services in my area to meet the community's needs and expectations."*

The mode for these questions is either 4 or 5, suggesting a strong positive opinion toward government policies and services consumed within the smart city. The standard deviation for Q5 is lower than both Q4 and Q12, which suggests that respondents are mainly content with communication policies regarding urban services but have varying opinions on new projects and services.

### $X_3$ Environmental Sustainability (Q6, Q7, Q15)

*"Q6 – Smart city projects in my city have positively impacted the local environment and green spaces."*

*"Q7 - Parks and green spaces are abundantly accessible in my city."*

*"Q15 – I do not intend to move from my city within the next five years."*

The mode for Q6 and Q15 are 5 and 4, respectively, suggesting that the environmental climate within the smart city is positively perceived. This is further supported by the mean and median for both questions, equating to 4 – Somewhat Agree.

Q7, however, has a neutral response and a mode of 1. With the standard deviation leaning to 1.45, there is a diverse opinion on the subject but favors a negative outlook, which may point to underlying issues with parks and recreational areas within the city.

### $X_4$ Socioeconomic (Q9, Q10, Q11, Q13, Q14)

*"Q9 – Income-controlled housing is accessible through most establishments in my city."*

*"Q10 - There is a noticeable homeless population across my city."*

*"Q11 - I get paid a higher salary for my job in my city than if I worked that same job elsewhere in the U.S."*

*"Q13 - Smart city services have improved my work or personal life."*

*"Q14 - The cost of living in my city is significantly more expensive than anywhere else in the U.S."*

The mean, mode, and median for Q9 and Q10 were all 3, indicating an almost perfect neutral response. The subject of homelessness and affordable rent may have been sensitive, resulting in a non-committal answer. The standard deviation is slightly higher for Q9 than Q10 but indicates varying opinions for these questions.

Q11, Q13, and Q14 have a mode of either 4 or 5, indicating a favorable view of social and economic factors within a smart city. The standard deviation is also, on average, 1.2 or below, supporting the opinion that most respondents favor the socioeconomic benefits of residing in a smart city.

## LINEAR REGRESSION ANALYSIS

Linear regression analysis was performed on the survey data to determine whether the statistical method was appropriate for predictive analysis. Each question in the study, apart from the first one, evaluated a specific independent variable.

| Question | Multiple R | R Square | Adjusted R Square | Significance f | Coefficients | P-value | Level of Significance |
|---|---|---|---|---|---|---|---|
| Question 2 | 0.12 | 0.013 | -0.007 | 0.42 | 0.12 | 0.42 | Not Significant |
| Question 3 | 0.06 | 0.004 | -0.02 | 0.68 | 0.07 | 0.68 | Not Significant |
| Question 8 | 0.24 | 0.056 | 0.03 | 0.10 | 0.20 | 0.09 | Not Significant |

*Table 4: Linear Regression Statistics for X1*

The linear regression statistics for smart technology, variable ($X_1$), summarized in Table 4, reveals no significant relationship with the dependent variable across all three questions. This is supported by a p-value higher than 0.05 for questions related to the $X_1$ variable. The low R Square values across the board suggest that other variables not included in this model might better explain the variability in the dependent variable.

| Question | Multiple R | R Square | Adjusted R Square | Significance f | Coefficients | P-value | Level of Significance |
|---|---|---|---|---|---|---|---|
| Question 4 | 0.09 | 0.007 | -0.014 | 0.56 | 0.09 | 0.56 | Not Significant |
| Question 5 | 0.06 | 0.004 | -0.02 | 0.66 | 0.08 | 0.66 | Not Significant |
| Question 12 | 0.12 | 0.014 | -0.007 | 0.42 | -0.13 | 0.42 | Not Significant |

*Table 5: Linear Regression Statistics for X2*

Singh, Wayne ws2398@nyu.edu

Table 5 summarizes regression statistics for the independent variable Government Policy ($X_2$) in relation to the (Y) variable, Urbanization patterns.

Questions 4, 5, and 12 show weak correlations with the dependent variable, as indicated by low Multiple R values. The R Square values suggest that the variables explain a very small portion of the variance in the dependent variable. Negative Adjusted R Square values for Q4 and Q12 imply that these predictors do not add value to the model over and above what would be expected by chance. Additionally, the coefficients are not statistically significant for any of the questions, as evidenced by the high p-values (p > 0.05), which implies that there is no clear evidence that Government Policy, as measured by these questions, has a significant impact on the dependent variable within the dataset.

| Question | Multiple R | R Square | Adjusted R Square | Significance f | Coefficients | P-value | Level of Significance |
|---|---|---|---|---|---|---|---|
| Question 6 | 0.10 | 0.011 | -0.01 | 0.47 | 0.12 | 0.47 | Not Significant |
| Question 7 | 0.03 | 0.001 | -0.02 | 0.82 | 0.03 | 0.82 | Not Significant |
| Question 15 | 0.05 | 0.002 | -0.02 | 0.74 | 0.05 | 0.74 | Not Significant |

*Table 6: Linear Regression Statistics for X3*

Table 6 summarizes the regression statistics for the independent variable Environmental Sustainability ($X_3$) in relation to the (Y) variable Urbanization patterns.

The low R Square values across the three questions imply that other variables not included in this analysis may better explain the variability in the dependent variable. The negative Adjusted R Square values suggest that including these particular predictors does not enhance the model's ability to explain the dependent variable. Although not statistically significant, the coefficients consistently indicate a positive direction of effect, which could merit further investigation, perhaps with a larger sample size, additional variables, or a different modeling approach.

| Question | Multiple R | R Square | Adjusted R Square | Significance f | Coefficients | P-value | Level of Significance |
|---|---|---|---|---|---|---|---|
| Question 9 | 0.08 | 0.006 | -0.02 | 0.60 | -0.06 | 0.60 | Not Significant |
| Question 10 | 0.06 | 0.004 | -0.02 | 0.68 | 0.05 | 0.68 | Not Significant |
| Question 11 | 0.12 | 0.02 | -0.01 | 0.40 | 0.11 | 0.39 | Not Significant |
| Question 13 | 0.06 | 0.01 | -0.01 | 0.60 | 0.07 | 0.60 | Not Significant |
| Question 14 | 0.06 | 0.004 | -0.02 | 0.65 | -0.07 | 0.66 | Not Significant |

*Table 7: Linear Regression Statistics for X4*

Table 7 summarizes the regression statistics for the Socioeconomic ($X_4$) independent variable in relation to the study's (Y) variable, Urbanization patterns. As indicated by a p-value greater than (0.05), the regression analysis suggests weak relationships with the dependent variable across Questions 9, 10, 11, 13, and 14. None of the R Square values are substantial enough to indicate a strong predictive relationship, and the negative Adjusted R Square values for each question point to a lack of model fit with the inclusion of these variables. The coefficients, both positive (Q10, Q11, Q13) and negative (Q9, Q14), are very small and therefore not statistically significant, which indicates that within this dataset, socioeconomic factors do not have a considerable influence on the dependent variable measured.

Hypothesis Testing

**Smart City Technology ($X_1$)**

|  | Q1 | Q2 |
|---|---|---|
| Mean | 3.92 | 3.82 |
| Variance | 1.25877551 | 1.293469388 |
| Observations | 50 | 50 |
| Hypothesized Mean Difference | 0 |  |
| df | 98 |  |
| t Stat | 0.44261266 |  |
| P(T<=t) one-tail | 0.32951066 |  |
| t Critical one-tail | 1.66055122 |  |
| P(T<=t) two-tail | 0.65902131 |  |
| t Critical two-tail | 1.98446745 |  |

*Table 8: Hypothesis Statistical Analysis for X1*

$H_0$: Smart City Technology significantly impacts urbanization patterns

$H_a$: Smart City Technology does not impact urbanization patterns

    From the data in Table 8, we can conclude that P(T<=t) two tail is greater than 0.05, and we therefore accept $H_0$ and reject $H_a$.

**Government Policy ($X_2$)**

|  | Q1 | Q5 |
|---|---|---|
| Mean | 3.92 | 3.72 |
| Variance | 1.25877551 | 1.34857143 |
| Observations | 50 | 50 |
| Hypothesized Mean Difference | 0 |  |
| df | 98 |  |
| t Stat | 0.87582147 |  |
| P(T<=t) one-tail | 0.19163437 |  |
| t Critical one-tail | 1.66055122 |  |
| P(T<=t) two-tail | 0.38326875 |  |
| t Critical two-tail | 1.98446745 |  |

*Table 9: Hypothesis Statistical Analysis for X2*

$H_0$: Government Policy directly impacts urbanization patterns

$H_a$: Government Policy does not impact urbanization patterns

From the data in Table 9, we can conclude that P(T<=t) two tail is greater than 0.05, and we therefore accept $H_0$ and reject $H_a$.

**Environment ($X_3$)**

|  | Q1 | Q15 |
|---|---|---|
| Mean | 3.92 | 3.56 |
| Variance | 1.25877551 | 1.47591837 |
| Observations | 50 | 50 |
| Hypothesized Mean Difference | 0 | |
| df | 97 | |
| t Stat | 1.539334999 | |
| P(T<=t) one-tail | 0.063489222 | |
| t Critical one-tail | 1.66071461 | |
| P(T<=t) two-tail | 0.126978445 | |
| t Critical two-tail | 1.984723186 | |

*Table 10: Hypothesis Statistical Analysis for X3*

$H_0$: Environment directly impacts urbanization patterns

$H_a$: Environment does not impact urbanization patterns

From the data in Table 10, we can conclude that P(T<=t) two tail is greater than 0.05, and we therefore accept $H_0$ and reject $H_a$.

**Socioeconomic ($X_4$)**

|  | Q1 | Q14 |
|---|---|---|
| Mean | 3.92 | 3.82 |
| Variance | 1.25877551 | 1.08938776 |
| Observations | 50 | 50 |
| Hypothesized Mean Difference | 0 |  |
| df | 97 |  |
| t Stat | 0.46144597 |  |
| P(T<=t) one-tail | 0.3227559 |  |
| t Critical one-tail | 1.66071461 |  |
| P(T<=t) two-tail | 0.6455118 |  |
| t Critical two-tail | 1.984723186 |  |

*Table 11: Hypothesis Statistical Analysis for X4*

$H_0$: Socioeconomic directly impacts urbanization patterns

$H_a$: Socioeconomic does not impact urbanization patterns

    From the data in Table 11, we can conclude that P(T<=t) two tail is greater than 0.05, and we therefore accept $H_0$ and reject $H_a$.

CORRELATION ANALYSIS

|  | Q1 | Q2 | Q3 | Q4 | Q5 | Q6 | Q7 | Q8 | Q9 | Q10 | Q11 | Q12 | Q13 | Q14 | Q15 |
|---|---|---|---|---|---|---|---|---|---|---|---|---|---|---|---|
| Q1 | 1 |  |  |  |  |  |  |  |  |  |  |  |  |  |  |
| Q2 | 0.116434874 | 1 |  |  |  |  |  |  |  |  |  |  |  |  |  |
| Q3 | 0.1192579 | 0.22931853 | 1 |  |  |  |  |  |  |  |  |  |  |  |  |
| Q4 | 0.143200309 | 0.04999891 | -0.05300645 | 1 |  |  |  |  |  |  |  |  |  |  |  |
| Q5 | 0.060774938 | -0.10074801 | -0.09735086 | -0.09223375 | 1 |  |  |  |  |  |  |  |  |  |  |
| Q6 | 0.037022087 | -0.04622341 | 0.03071883 | -0.20858568 | 0.20902089 | 1 |  |  |  |  |  |  |  |  |  |
| Q7 | 0.03351314 | 0.05279833 | 0.00590271 | -0.17912049 | -0.04977539 | -0.2503044 | 1 |  |  |  |  |  |  |  |  |
| Q8 | 0.237367165 | -0.02363104 | -0.04229521 | 0.20756406 | -0.13096965 | 0.01796806 | -0.11085691 | 1 |  |  |  |  |  |  |  |
| Q9 | -0.07665905 | -0.03386841 | -0.04301191 | 0.04356231 | 0.26898911 | -0.04027285 | 0.05876457 | 0.00829113 | 1 |  |  |  |  |  |  |
| Q10 | 0.059861379 | 0.32948324 | -0.13644089 | -0.07773011 | 0.08756189 | 0.24652172 | -0.14800683 | 0.14371526 | -0.18176119 | 1 |  |  |  |  |  |
| Q11 | 0.123000741 | 0.19280217 | 0.24744227 | -0.14891854 | 0.05884625 | -0.06013323 | 0.1974954 | 0.131059 | 0.0270175 | 0.03468971 | 1 |  |  |  |  |
| Q12 | -0.175328622 | -0.00992401 | 0.01837242 | -0.0677462 | 0.26102855 | -0.00512765 | 0.08645941 | 0.11340687 | 0.23762276 | -0.01487721 | -0.10876768 | 1 |  |  |  |
| Q13 | 0.075977052 | 0.36413292 | -0.16355109 | 0.05046566 | -0.09710133 | 0.04482531 | 0.02399423 | 0.04570236 | 0.19393065 | 0.06038261 | -0.09655633 | -0.17499676 | 1 |  |  |
| Q14 | -0.064830814 | -0.07942854 | 0.09563187 | -0.00317087 | 0.1596188 | 0.12063284 | -0.07688812 | -0.08427127 | 0.07684265 | -0.01683852 | -0.02828703 | 0.05097883 | -0.14983647 | 1 |  |
| Q15 | 0.048511379 | -0.30958931 | -0.21143797 | 0.04011998 | 0.09894455 | 0.26818076 | -0.28593936 | 0.08698077 | -0.12334943 | 0.06509932 | -0.01692969 | -0.04777915 | -0.13204428 | 0.25815817 | 1 |

*Table 12: Correlation Analysis- Survey Data*

    Table 12 illustrates the correlations between the dependent variable and the independent variables. It also shows the relationship among independent variables with possible multicollinearity.

The strongest correlation with the dependent variable is Question 8 at (24%), representing smart technology mobility. Questions 10 and 13 evaluate socioeconomic factors of housing and work satisfaction, respectively, and stand out at (33%) and (36%) in correlation to Question 2 (technology that contributes to sustainable living). Therefore, the relationship between technology's role in supporting affordable housing and improving work-life balance may be an appropriate topic for further studies.

## FINDINGS

The findings underscore the complexity and dynamism of defining and implementing smart city technologies, which may contribute to the observed discrepancies between quantitative and qualitative results. The rapid acceleration of technology implementation in urban areas, coupled with a lack of uniform taxonomy and awareness among respondents, can pose significant challenges in quantitatively measuring the impact of such technologies on urbanization. Respondents may not fully understand or be aware of how smart city initiatives are being implemented, complicating efforts to capture their actual effects through survey-based methods.

These findings highlight the importance of adopting a comprehensive, mixed-methods approach when studying the dynamics of urbanization in the context of smart cities. Policymakers and urban planners should consider the measurable impacts captured through quantitative methods and qualitative research's subjective, nuanced insights. This dual approach offers a complete picture of how various factors contribute to urbanization patterns, facilitating more informed and effective urban planning and policymaking.

## Limitations

This study encountered several limitations that influenced the findings and interpretation. The primary limitation was the relatively small sample size (n=50), which may have restricted the ability to detect significant effects of the independent variables on urbanization patterns. Consequently, the limited sample also confined the scope of the study to two of the fourteen possible cities undertaking urban smart technology initiatives. These smart cities have progressed at different rates and would, therefore, collectively offer a more accurate reflection of smart city influence on urbanization.

Additionally, the rapid evolution of smart city technologies and the variability in public understanding of these concepts or whether they exist likely introduced biases or misinterpretations in the quantitative data collected.

## Recommendations

Future research should consider a more extensive and diverse sample, increasing the number of participants to at least 3,500 across more U.S. cities. This expansion would improve the representative data and enhance the statistical power of the analyses, allowing for a finer detection of independent variable effects.

The study aimed to assess the impact of smart city technologies, government policies, environmental sustainability, and socioeconomic factors on urbanization patterns in the United States. While the quantitative findings did not show significant impacts, qualitative data suggested that these factors play an essential role. This discrepancy highlights the complexity of urban systems and the potential for qualitative insights to capture nuances that quantitative methods might miss. Thus, the study partially achieved its objective by illuminating these complexities and providing a dual perspective understanding of the influences on urbanization.

While this study's limitations impacted the outcomes, the insights gained are invaluable for understanding the complex dynamics evolving within smart city ecosystems. By addressing these limitations and implementing the recommended strategies, future research can contribute more effectively to develop smart cities that are equitable, sustainable, and responsive to the needs of their inhabitants.

**CONCLUSION**

This study systematically explored the impact of smart city technologies, government policies, environmental sustainability measures, and socioeconomic factors on urbanization patterns in the United States. The research utilized a mixed-methods approach that integrated quantitative and qualitative data to provide a multifaceted understanding of how these variables influence urban development, summarized in Table 13 below.

| Variables | Research Question | Quantitative Data (Qt.D) | Qualitative Data (Q1.D) | Agreement with Qt.D |
|---|---|---|---|---|
| Smart City Technology $X_1$ | Does smart city technology influence urbanization patterns in the United States | $X_1$ does not have a significant impact on urbanization patterns. P value> 0.05 | Smart city technology plays a pivotal role in urban transformation by optimizing city functions and enhancing public services, leading to increased urbanization concentration (Kummitha, 2018) | No |
| Government Policy $X_2$ | Does government policy influence urbanization | $X_2$ does not have a significant impact on | Government policies significantly shape the development | No |

| | patterns in the United States | urbanization patterns. P value> 0.05 | and integration of smart technologies into urban environments, thereby influencing urbanization patterns (Mohanty & Kumar, 2021b) | |
|---|---|---|---|---|
| Environmental Sustainability $X_3$ | Does environmental sustainability influence urbanization patterns in the United States | $X_3$ does not have a significant impact on urbanization patterns. P value> 0.05 | Environmental sustainability measures are essential in smart city strategies, mitigating adverse ecological impacts and improving living conditions, thereby influencing urbanization(Chivas et al., 2023) | No |
| Socioeconomic Factors $X_4$ | Do socioeconomic factors influence urbanization patterns in the United States | $X_4$ does not have a significant impact on urbanization patterns. P value> 0.05 | Socioeconomic dynamics such as income levels, employment, and social inequality are closely linked to urban reform cycles and significantly impact the development and attractiveness of smart cities (Angelo & Vormann, 2018) | No |

*Table 13: Conclusion Table*

As the data in Table 13 confirms, the quantitative analysis revealed that none of the independent variables—smart city technology ($X_1$), government policy ($X_2$), environmental sustainability ($X_3$), and socioeconomic factors ($X_4$)—had a statistically significant impact on urbanization patterns, as indicated by P-values greater than 0.05 across the categories. This suggests that within the scope of this study, the direct effects of these variables on urbanization may be limited or influenced by other factors not captured in the survey data.

However, the qualitative data provided a contrasting perspective, suggesting the nuanced impact of these variables on urbanization. For instance, qualitative insights indicated that smart city technologies play a crucial role in urban transformation by optimizing city functions and enhancing public services, potentially leading to increased urban concentration. Similarly, government policies significantly shaped the development and integration of smart technologies into urban environments, influencing urbanization patterns. Environmental sustainability measures were highlighted as essential in mitigating adverse ecological impacts and improving living conditions, affecting urbanization. Lastly, socioeconomic dynamics such as income levels, employment, and social inequality were linked to urban reform cycles and significantly influenced the development and attractiveness of smart cities.